\documentclass[conference]{IEEEtran}

\usepackage{cite}
\usepackage{graphicx}
\usepackage{xcolor}
\usepackage{hyperref}
\hypersetup{colorlinks=true,linkcolor=black,citecolor=blue,filecolor=black,urlcolor=blue}
\graphicspath{{figures}}
\usepackage{caption}
\usepackage{subcaption}
\usepackage{listings}
\renewcommand{\lstlistingname}{Listing}
\usepackage[binary-units=true]{siunitx}
\usepackage{csvsimple}
\usepackage{booktabs}
\usepackage{amsmath}

\usepackage{color,soul}

\title{An Analysis of Performance Bottlenecks in MRI Pre-Processing}
\author{Mathieu Dugr\'e*, Yohan Chatelain, Tristan Glatard
\\Department of Computer Science and Software Engineering, Concordia University, Montreal, QC, Canada
\\\{mathieu.dugre, yohan.chatelain, tristan.glatard\}@concordia.ca}

\begin{document}
\maketitle

\begin{abstract}
	Magnetic Resonance Image (MRI) pre-processing is a critical step for neuroimaging analysis. However, the computational cost of MRI pre-processing pipelines is a major bottleneck for large cohort studies and some clinical applications. While High-Performance Computing (HPC) and, more recently, Deep Learning have been adopted to accelerate the computations, these techniques require costly hardware and are not accessible to all researchers. Therefore, it is important to understand the performance bottlenecks of MRI pre-processing pipelines to improve their performance. Using Intel VTune profiler, we characterized the bottlenecks of several commonly used MRI-preprocessing pipelines from the ANTs, FSL, and FreeSurfer toolboxes. We found that few functions contributed to most of the CPU time, and that linear interpolation was the largest contributor. Data access was also  a substantial bottleneck. We identified a bug in the ITK library that impacts the performance of ANTs pipeline in single-precision and a potential issue with the OpenMP scaling in FreeSurfer recon-all. Our results provide a reference for future efforts to optimize MRI pre-processing pipelines.
\end{abstract}

\section{Introduction}
Pre-processing of Magnetic Resonance Imaging (MRI) data is a critical step in neuroimaging. Although established pre-processing pipelines exist, they commonly require extensive amounts of computation and produce large volumes of output and intermediate data. Such requirements create challenges when studying large neuroimaging cohorts, and they limit clinical applicability when timely data analysis is needed. Therefore, the neuroimaging community is constantly seeking for ways to improve the computational performance of MRI pre-processing pipelines. In this paper, we characterize the computational cost of several commonly adopted MRI pre-processing pipelines, providing a reference benchmark for future efforts to optimize MRI pre-processing pipelines.

MRI is a standard tool used by neuroscientists to perform clinical diagnosis and for researchers to develop a better understanding of the brain. Three main MRI modalities exist: structural MRI (sMRI), functional MRI (fMRI), and diffusion MRI (dMRI). While other modalities such as electroencephalography (EEG), computed tomography (CT), and positron emission tomography (PET) exist, we focus on MRI for its broad adoption in research and non-invasiveness. Moreover, MRI data analysis is challenging due to computationally expensive pre-processing and large output and intermediate data size.

Neuroscientists developed various toolboxes to tackle the challenging task of pre-processing MRI data. We focus on the commonly accepted fMRIPrep~\cite{Esteban2019-bl} pipeline as it a comprehensive pre-processing application for both sMRI and fMRI. fMRIPrep combines several pipelines to produce a complete pre-processing pipeline. We study fMRIPrep's sub-pipelines separately, to provide a finer grained analysis of the performance bottlenecks.

To profile the pipelines, we use the Intel VTune tool~\cite{vtune-profiler}, a multi-language profiler with low performance overhead that provides runtime information at the level of functions. We profile the pipeline in single-threaded mode and using 32 threads to account for parallel executions. We aggregate subjects' results for each pipeline to obtain an average performance profile. 

\section{Materials \& Methods}
\subsection{Pipelines}
fMRIPrep is a commonly used fMRI and sMRI pre-processing pipeline for neuroimaging, built on top of several well-known neuroimaging toolboxes such as Advanced Normalization Tools (ANTs)~\cite{Avants2020-xx}, FMRIB Software Library (FSL)~\cite{Jenkinson2012-tq}, FreeSurfer~\cite{Fischl2012-cx} and Analysis of Functional NeuroImages (AFNI)~\cite{Cox1996-nk}. Although processing steps may vary based on the type of input data, the sMRI pipeline generally uses ANTs BrainExtraction~\cite{Tustison2010-gg,Avants2008-ea} to perform intensity correction and skull-stripping of a T1 weighted image, FSL FAST~\cite{Zhang2001-hx} to segment the tissues of the extracted brain, and ANTs registrationSyN~\cite{Avants2008-ea} to register the segmented brain. When enabled, FreeSurfer recon-all~\cite{Dale1999-wu} is used for surface reconstruction. We benchmarked a recon-all command similar to  fMRIPrep's, which (1) performs autorecon1 without skull-stripping, (2) imports an external skull-strip from ANTs brainExtraction, and (3) resumes reconstruction with autorecon2 and autorecon3. The fMRI pipeline generally performs motion correction with FSL MCFLIRT~\cite{Jenkinson2002-od}, slice timing correction with AFNI 3dtshift~\cite{Cox1996-nk}, and co-registration with FSL FLIRT~\cite{Jenkinson2002-od,Jenkinson2001-eu,Greve2009-dw}.

We profiled the entire fMRIPrep pipeline to obtain a coarse view of the pipeline's performance bottleneck. Additionally, to characterize the performance at a finer grain, we profiled the default sub-pipelines of fMRIPrep listed above. While we used the same sub-pipeline and order as in fMRIPrep, we used the default parameters for most of the sub-pipelines which may vary from the default parameters of fMRIPrep. We omitted some sub-pipelines of fMRIPrep as they were not compatible with our dataset. For example, slice-timing correction was already performed in our data.

ANTs brainExtraction and registrationSyN are available in single or double precision, leveraging the Insight Segmentation and Registration Toolkit (ITK)~\cite{Yoo2002-ve}. We profiled both versions to understand the performance benefits of using reduced precision in the pipelines.

\subsection{Data}
To ensure that our performance profiles are inclusive of different populations, we used data with a wide range of age and equal distribution of sex. We used the OpenNeuro ds004513 v1.0.2 dataset~\cite{ds004513:1.0.2}. The anatomical and functional data from 20 health individuals was acquired from two different cohorts. The within-subject cohort has nine participants (mean age=43~yrs, std=7~yrs; 4 females) with two sessions: eyes open and eyes closed. The replication cohort has eleven participants (mean age=27~yrs, std=5~yrs; 6 females) with only the eyes open session. Each subject has an anatomical image with resolution of 1mm and a dimension of $160 \times 240 \times 256$ voxels (approx. \SI{12}{\mebi\byte}). Additionally, the subjects have 300 functional images taken at 2 seconds interval with a resolution of 3mm and a dimension of $64 \times 64 \times 35$ voxels (approx. \SI{50}{\mebi\byte}).
\subsection{Profiling}
Profiling MRI pre-processing pipelines raises a few challenges. Neuroimaging pipelines use several programming languages including \textit{C}, \textit{C++}, \textit{Fortran}, \textit{Matlab}, and \textit{Python}, with several pipelines utilizing a combination of multiple languages. Pipelines are generally complex and computationally expensive.

The Intel VTune profiler addresses these challenges by offering multi-language support, low performance overhead, and information of functions at runtime level. However, common profiling challenges remain. Profilers require source code to be compiled with debug symbols to report interpretable information about the different functions and modules. Moreover, different input data can substantially impact the performance results due to conditional branching in the pipeline and convergence thresholds. We discuss our approach to these challenges in the next section.
		
First, to obtain human-readable information on function and module names in VTune, we re-compiled each pipeline with debug information in a Docker image. For use on HPC systems, we created Apptainer~\cite{Kurtzer2017-bu} images using Docker2Singularity\footnote{\href{https://github.com/singularityhub/docker2singularity}{https://github.com/singularityhub/docker2singularity}}. \lstlistingname~\ref{lst:vtune_example} demonstrates the profiling of a pipeline by mounting the VTune binary to the Apptainer image during execution.

\begin{minipage}{0.9\linewidth}
\begin{lstlisting}[
		language={sh},
		caption={Profiling a pipeline within an Apptainer container, using VTune profiler.},
		captionpos=b,
		label=lst:vtune_example,
		numbers=left,
		frame=single,
		basicstyle=\footnotesize,
		numbersep=5pt,
		numberstyle=\tiny\color{gray},
		rulecolor=\color{black},
		tabsize=2,
	]
#!/bin/sh
singularity exec --cleanenv \
    -B ~/intel/oneapi/vtune/latest/:/vtune \
    -B ${SLURM_TMPDIR}:/data \
    -B ${PROJECT_DIR}:${PROJECT_DIR} \
    ${SIF_IMG} \
    /vtune/bin64/vtune \
    -collect hpc-performance \
    -knob enable-stack-collection=true \
    -knob analyze-openmp=true \
    -result-dir ${PROFILING_DIR} \
    <script.sh>
\end{lstlisting}
\end{minipage}
			
We profiled the pipeline using two threading configuration: single-threaded and 32 threads. The single-threaded approach simplified the analysis and limited the potential I/O overhead on the profiling. 
The 32-thread approach used all available CPU cores from a compute node of our infrastructure to study the multi-threading performance of the pipeline. This configuration provided a more realistic usage of the pipeline.
			
Before profiling each pipeline, we transferred the input data from the shared file system to the compute node. After profiling, we transferred back the output to the shared file system. This limited profiling variability due to I/O contingency.
			
The data generated by the VTune profiler was written to a shared file system, to simplify post-processing. In principle, this could have increased the profiling overhead. However, in practice there was no impact since the data generated by the VTune profiler was very small and was not written on the same file system as data produced by the pipelines.
			 
\subsection{Metrics Definition}
VTune provides several metrics to characterize the performance of a pipeline.  The makespan is the elapsed time from the start to the end of the pipeline. The CPU time refers to the total time spent by the CPU, across all cores, to execute the pipeline. 

The memory bound metrics counts the number of cycles where the CPU is starved while waiting for in-flight memory demand load. L1 bound occurs when a demand load stalls CPU cores although the data already reside in L1 cache. Similarly, L2, L3, and DRAM bound occurs when a demand load stalls the CPU and there is a cache miss in the previous level of cache, but the data reside in the L2, L3, and DRAM respectively. The store bound occurs when a store operation stalls the CPU. Supplementary Figure~\ref{fig:memory-metrics} shows the monitoring events and equation to derive the metrics defining the memory bound characterization~\cite{Intel2006-lc}. We use the convention from~\cite{Kukunas2015-jd} for the equations.
			
The difference between the memory bound value and the sum of the other metrics represent the percentage of CPU starvation while waiting from data absent from all cache level, that is, when the data is being fetched from disk.
			
\subsection{Infrastructure}
We used the \textit{Slashbin cluster} at Concordia University. The compute nodes are configured with two 16-core Intel(R) Xeon(R) Gold 6130 CPU @ \SI{2.10}{\giga\hertz}, \SI{250}{\gibi\byte} of RAM, \SI{126}{\gibi\byte} of tmpfs, six \SI{447}{\gibi\byte} SSDs with the XFS file system (no RAID enabled), Rocky~Linux~8.9 and Linux kernel \textit{4.18.0-477.10.1.el8\_lustre.x86\_64}.
			
\section{Results \& Discussion}
In this section we present the aggregated profiling data obtained with VTune. First, we present a performance overview of the pipelines. We then analysis specific hotspots in the pipelines. Unless specified explicitly, we only present the profile data from using 1-thread since the results from 32-threads are similar.

We aggregate the profiling results of each pipeline across all subjects to obtain an average performance profile. Consistently with~\cite{Kukunas2015-jd}, we hypothesized that the Pareto principle applied to the performance of MRI pre-processing pipelines, we therefore only studied the functions with the largest contribution to CPU time, up to 80\% cumulative total runtime of the pipeline. This heuristic allowed us to focus our efforts on the performance critical sections of a pipeline. We report the mean and standard deviation of the execution time for those functions.
			
\subsection{CPU time distribution across functions was long-tailed}
\begin{figure}[ht]
	\centering
	\includegraphics[width=\linewidth]{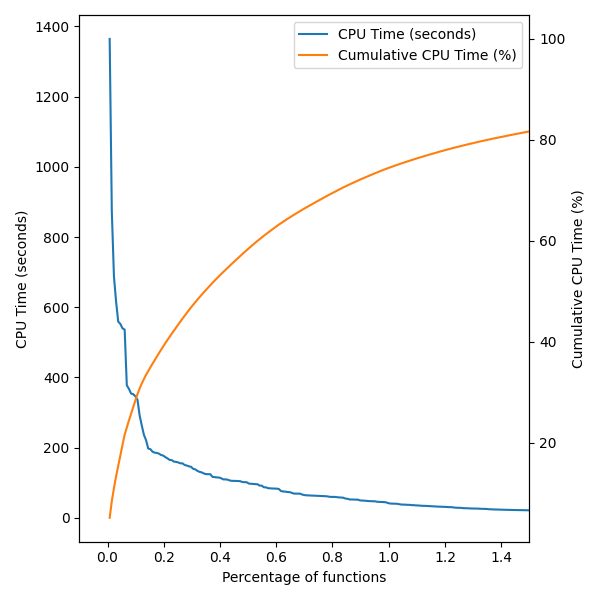}
	\caption{Distribution of the functions' CPU time. The left y-axis shows the average total CPU Time spent in a function, while the right y-axis show the cumulative CPU time percentage. The x-axis is the percentage of functions ordered by decreasing CPU time. The data includes all functions from all pipelines. }
	\label{fig:long-tail-distribution}
\end{figure}

\begin{figure*}[ht]
	\centering
	\includegraphics[width=0.66\columnwidth]{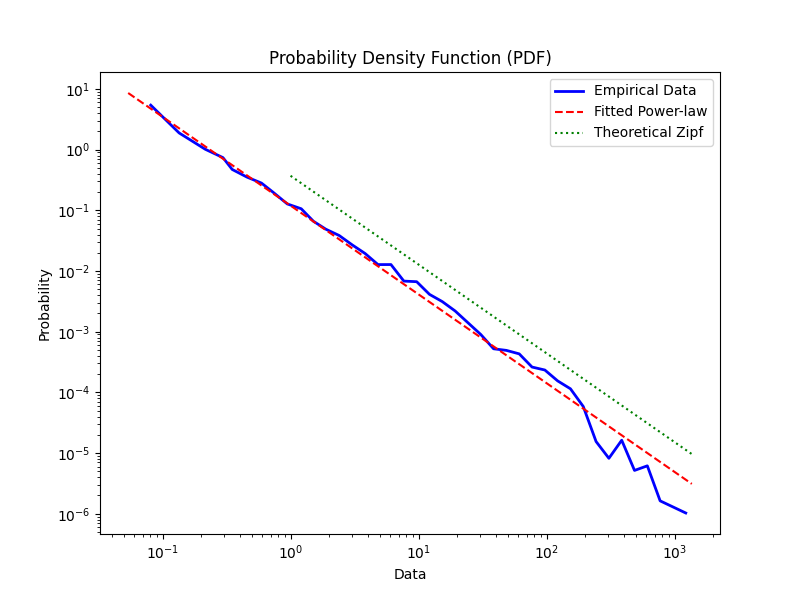}
	\caption{Distribution of the functions' CPU time compare to a Zipf's law distribution with $\alpha=1.46$. The y-axis shows the average CPU time for a function. The x-axis shows the percentage of functions ordered by decreasing CPU time. The data includes all functions from all pipelines.}
	\label{fig:zips-law}
\end{figure*}

Figure~\ref{fig:long-tail-distribution} shows that 80\% of the total CPU time was spent in less than 1.5\% of the function across all pipelines. While consistent with the Pareto principle, the distribution of the functions' CPU time did not follow a Pareto distribution. Using the \textit{powerlaw} Python package~\cite{Alstott2014-gr}, we found that the data fit a Zipf law with exponent $\alpha=1.46$ (Supplementary Figure~\ref{fig:zips-law}). Therefore, the CPU time of a function was approximately inversely proportional to its rank. Thus, future efforts to optimize this minimal set of functions could result in significant performance improvements.

\subsection{Main Bottleneck: Linear Interpolation}
Table~\ref{extab:interpolation} shows that interpolation was a critical part for all pipelines except FreeSurfer recon-all, contributing between 32.20\% to 62.70\% of the total CPU time. Overall, the number of functions using interpolation was low, with less than 20 in each pipeline: few interpolation functions contributed to a large portion of the CPU time. While each interpolation had low computational cost, the large amount of operations resulted in a significant bottleneck. Therefore, optimizing interpolation would bring substantial benefits to these pipelines.

\csvnames{csvcol}{1=\pipeline, 2=\nfunc, 3=\cputime}
\csvstyle{interpolation}{
	tabular = |r|c|c|,
	table head = \hline Pipeline & \# of functions & \% CPU Time \\
	& with interpolation & spent in interpolation\\\hline\hline,
	late after line = \\\hline,
	respect all,
	csvcol
}
\begin{table*}[ht]
	\centering
	\begin{filecontents*}{tables/interpolation.csv}
pipeline,# of functions with interpolation,% CPU Time with interpolation
ants/brainExtraction,17,38.52
ants/brainExtraction-fp,17,38.37
ants/registrationSyN,7,43.67
ants/registrationSyN-fp,8,47.14
fsl/fast,1,42.97
fsl/mcflirt,3,32.90
fsl/flirt,4,62.07
freesurfer/reconall,10,3.71
\end{filecontents*}
\csvreader[interpolation]{tables/interpolation.csv}{}
	{\pipeline & \nfunc & \tablenum[round-precision=2]{\cputime}}
	\caption{Contribution of interpolation to the pipelines' total CPU time. The percentage is the average sum of CPU time of functions using interpolation. The data includes all functions from all pipelines.}
	\label{extab:interpolation}
\end{table*}

\subsection{Impact of Memory Bounded Functions}
Table~\ref{extab:memory-bound} shows that fetching data from disk was the main cause for starving the CPU. Next, the L1 bound had the highest cache-level contribution to the memory bound of the pipelines. ANTs pipelines had the highest L1 bound, followed by FreeSurfer recon-all and FSL MCFLIRT. L2 and L3 bound was low for all pipelines, although higher for FreeSurfer recon-all. The DRAM bound was high for both FSL FLIRT and FreeSurfer recon-all, followed by ANTs brainExtraction in single and double precision. Store bound was low for all pipelines. We note an interesting relationship between the low L1 bound and the high DRAM bound of FreeSurfer recon-all and FSL FLIRT, compared to the other pipelines.

Optimizing the data access of the pipeline could reduce the number of data load. This would potentially have a large impact on the performance of the pipelines. Alternatively, the use of reduced precision could reduce the memory footprint, thus the memory bound. To reduce the memory bound impact of fetching data from disk, it could be possible to convert and store the data to a lower precision before pre-processing. In both cases, the impact of reduced precision on the accuracy of the pipelines should be studied.
			
\csvnames{csvcol}{1=\pipeline, 2=\mem, 3=\la, 4=\lb, 5=\lc, 6=\dram, 7=\store, 8=\disk}
\csvstyle{memory_bound}{
	tabular = |r|c|c|c|c|c|c|c|,
	table head = \hline Pipeline &  \% Memory Bound &  \% L1 Bound &  \% L2 Bound &  \% L3 Bound & \% DRAM Bound & \% Store Bound & \% Disk Bound\\\hline\hline,
	late after line = \\\hline,
	respect all,
	csvcol
}
\begin{table*}[ht]
	\centering
	\begin{filecontents*}{tables/memory_bound.csv}
pipeline, Memory Bound (%), L1 Bound(%), L2 Bound (%), L3 Bound (%), DRAM Bound (%), Store Bound (%), Disk Bound (%)
ants/brainExtraction,22.61,7.08,0.47,0.81,5.08,0.23,8.94
ants/brainExtraction-fp,20.59,9.11,0.24,0.39,2.48,0.12,8.24
ants/registrationSyN,16.01,6.61,0.79,0.33,0.70,0.44,7.14
ants/registrationSyN-fp,17.97,9.65,0.30,0.08,0.30,0.16,7.48
fsl/fast,9.87,2.87,0.40,0.51,1.61,1.67,2.81
fsl/mcflirt,6.79,3.57,1.96,0.19,0.16,0.10,0.81
fsl/flirt,30.36,1.61,0.42,2.37,15.73,0.59,9.64
freesurfer/reconall,34.07,4.11,2.50,4.69,13.83,1.23,7.72
\end{filecontents*}
\csvreader[memory_bound]{tables/memory_bound.csv}{}
	{\pipeline & \tablenum[round-precision=2]{\mem} & \tablenum[round-precision=2]{\la} & \tablenum[round-precision=2]{\lb} & \tablenum[round-precision=2]{\lc} & \tablenum[round-precision=2]{\dram} & \tablenum[round-precision=2]{\store} & \tablenum[round-precision=2]{\disk}}
	\caption{Impact of data load on stalled CPU cycles. For each metric (memory, L1, L2, L3, DRAM, and store), reported values are summations weighted by function CPU time, averaged over n=20 subjects, and obtained in single-threaded mode. They represent the percentage of the total CPU time stalled for the metric.}
	\label{extab:memory-bound}
\end{table*}			
						
\subsection{ANTs: Single vs. Double Precision}
\begin{figure}[ht]
	\includegraphics[width=\linewidth]{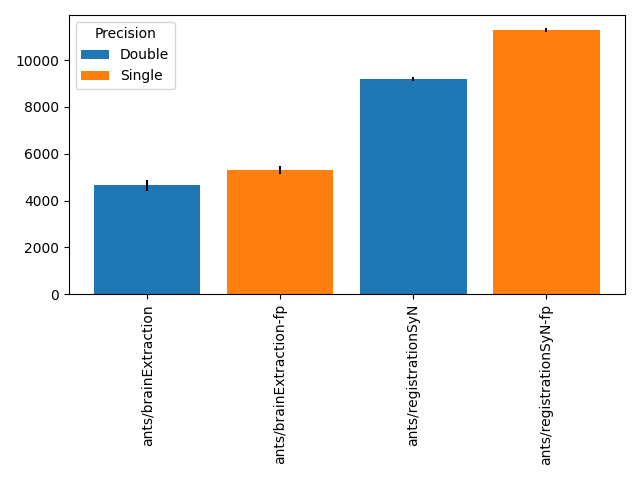}
	\caption{Comparison of makespan between double (blue) and single (orange) precision for ANTs brainExtraction and ANTs registrationSyN.}
	\label{fig:makespan-ants}
\end{figure}

Figure~\ref{fig:makespan-ants} shows the makespan of ANTs brainExtraction and ANTs registrationSyN in double and single precision. For both pipelines, the makespan was significantly higher in single precision than in double precision. This was unexpected, as the floating-point arithmetic operations are supposed to be faster in single precision than in double precision~\cite{Wang2018-jv}.

\begin{figure}
	\includegraphics[width=\linewidth]{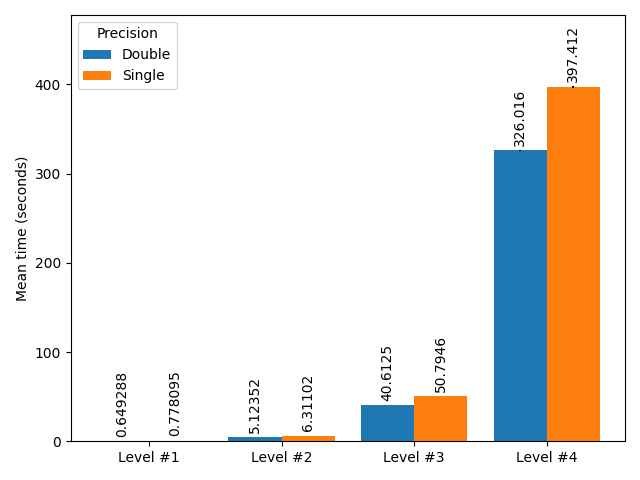}
	\caption{Average time per iteration for ANTs registrationSyN in double and single precision. Only the SyN Registration stage is shown as the two earlier stages were near zero time. The error bars show the standard deviation across n=20 subjects.}
	\label{fig:mean-time-per-iteration-ants}
\end{figure}

ANTs registrationSyN is a three-stage process (Euler transform, affine transform, and SyN registration), where each stage is successively performed at 4 resolution levels with a specified number of maximum iterations. We focused on the SyN registration stage since it was substantially longer to execute than the other stages. The number of iterations between both versions of ANTs registrationSyN was similar. In fact, for the last two levels, the maximum specified number of iterations was reached across all but one execution. However, the single precision version of ANTs registrationSyN took approximately 20-25\% longer per iteration than the double precision version for the last registration stage (Figure~\ref{fig:mean-time-per-iteration-ants}). Therefore, the slowdown was not due to a slower convergence of the algorithm, but rather to an increased execution time for each iteration.

After further analysis of the ANTs registrationSyN single-precision pipeline, we found that the use of templates in ITK's C++ code\footnote{\href{https://github.com/InsightSoftwareConsortium/ITK/blob/d9c585d96359bf304ad3047148cee81bf27ac0c1/Modules/Core/ImageFunction/include/itkVectorInterpolateImageFunction.h\#L46-L48}{itkVectorInterpolateImageFunction.h}} forces the output of the interpolation function to be in double precision, with \textit{EvaluateAtContinuousIndex} contributing most of the CPU time. Therefore, the pipeline wastes CPU cycles to convert the input data to double-precision in ITK, and potentially to cast back to single-precision in ANTs. We think this bug causes the slow-down. We tried recompiling a version of ITK to fix this issue, but we were unsuccessful due to the size and complexity of the code base. This example shows that the use of reduced precision is not trivial and requires a deep understanding of the code base. Moreover, benchmarks should be performed to ensure that the reduced precision does not impact the performance or accuracy of the pipeline. We reported this issue in the ITK GitHub repository\footnote{\href{https://github.com/InsightSoftwareConsortium/ITK/issues/4593}{https://github.com/InsightSoftwareConsortium/ITK/issues/4593}}.

\subsection{FreeSurfer: Thread-Synchronization}
\begin{figure*}
	\centering
	\begin{subfigure}[t]{0.49\textwidth}
		\caption{Single-threaded}
		\label{subfig:hotspots-freesurfer-reconall-1thread}
		\includegraphics[width=\textwidth]{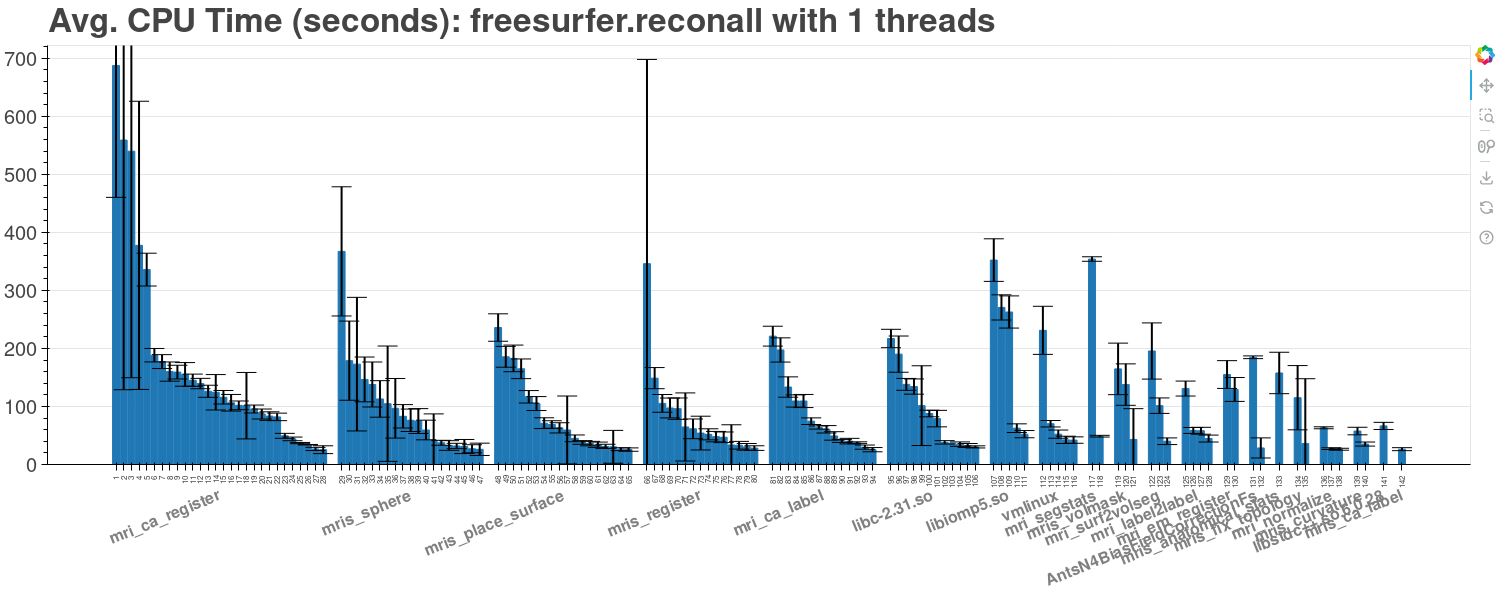}
	\end{subfigure}
	\begin{subfigure}[t]{0.49\textwidth}
		\caption{Multi-threaded (32 threads)}
		\label{subfig:hotspots-freesurfer-reconall-32threads}
		\includegraphics[width=\textwidth]{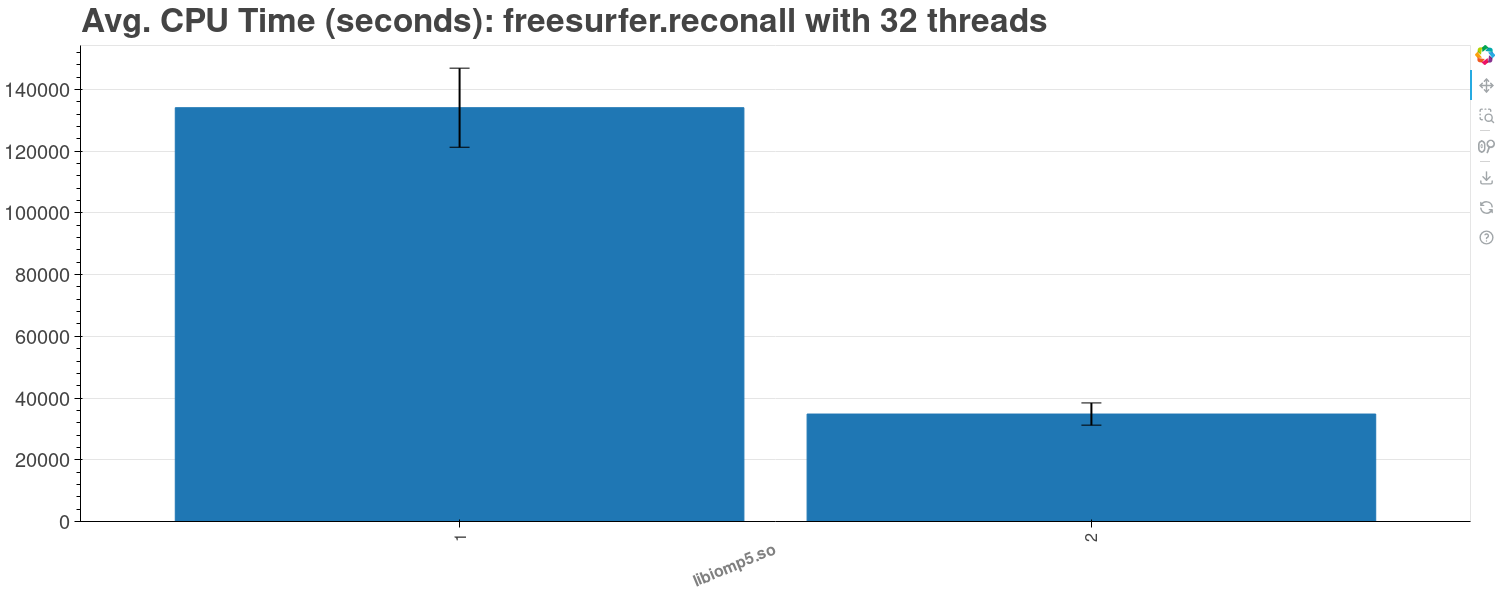}
	\end{subfigure}
	\caption{FreeSurfer recon-all analysis. The y-axis shows the average CPU time spent in each function, with error bars showing the standard deviation across n=20 subjects. The x-axis shows the function ordered by decreasing CPU time grouped by module. We omitted function names for clarity. The function ID are dependent to each plot. Supplementary materials show the mapping of the function ID to the function name for each plot.}
	\label{fig:hotspots-freesurfer-reconall}
\end{figure*}

Figure~\ref{fig:hotspots-freesurfer-reconall} shows a significant difference in the pipeline profile between single-threaded and multi-threaded executions. In the multi-threaded executions (Figure~\ref{subfig:hotspots-freesurfer-reconall-32threads}), the OpenMP multithreading library was a major bottleneck for the pipeline, accounting for at least 76\% of the pipeline runtime. Moreover, the y-axis scale is multi folds larger than for the single-threaded execution, indicating that the multi-threaded execution was significantly slower than the single-threaded execution.
					
\begin{figure}
	\includegraphics[width=\linewidth]{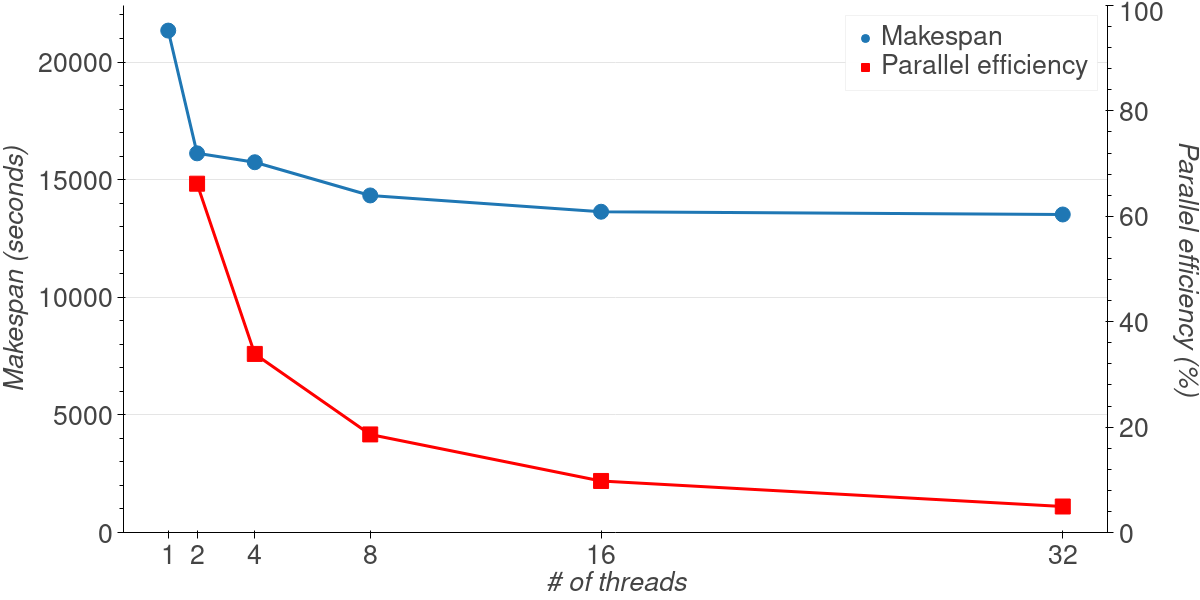}
	\caption{Makespan and parallel efficiency of FreeSurfer recon-all while varying the number of threads from 1 to 32. The left y-axis shows the makespan in seconds, while the right y-axis shows the parallel efficiency in percent. The log-scaled x-axis shows the number of threads.}
	\label{fig:freesurfer-threading}
\end{figure}

While the makespan for FreeSurfer recon-all decreased with an increase in the number of threads, the benefits were limited (Figure~\ref{fig:freesurfer-threading}). The parallel efficiency decreased from 66\% with two threads down to 5\% with 32 threads. While Amdhal's law limits the maximum parallel efficiency of the pipeline, Figure~\ref{subfig:hotspots-freesurfer-reconall-32threads} shows that most of the CPU time is spent in libiomp, which suggests that thread synchronization had a larger impact on parallel efficiency.	

OpenMP offers multiple scheduling policies to assign chunks to threads, including \textit{static} scheduling that assigns chunks to threads in round-robin fashion, and \textit{dynamic} which assign chunks to threads as they become available for work. In the codebase, 83 out of 92 tasks use the \textit{static} scheduling policy. This could lead to a load imbalance between the threads, and thus a decrease in parallel efficiency. We speculate that \textit{dynamic} scheduling could improve the parallel efficiency, by assigning chunks based on the current load of the threads. Unfortunately, we failed to recompile FreeSurfer having changed the \textit{static} OpenMP scheduling to \textit{dynamic}. This remains challenging as the codebase is large and complex, and the change in scheduling policy could lead to unexpected bugs.
We opened an issue on FreeSurfer GitHub repository to report this problem\footnote{\href{https://github.com/freesurfer/freesurfer/issues/1199}{https://github.com/freesurfer/freesurfer/issues/1199}}.
			
Future work should study the different configuration of OpenMP scheduling in FreeSurfer recon-all to improve the parallel efficiency. We believe that significant performance improvements could be achieved by optimizing the thread synchronization. This would lead to faster runtime time and higher CPU usage ration in HPC allocations.

\subsection{Experiencing using VTune}
The VTune profiler provided a rich and extensive amount of data for our analysis, which would have been difficult to retrieve otherwise. Thanks to its low performance overhead, we were able to profile the pipeline with minimal impact on the runtime, allowing the use of more subjects to obtain a more accurate performance profile. The documentation provided by VTune is extensive and well-documented, which helped us to understand the different metrics and events available.

VTune requires a finalization step to query the data and generate a report. We found this step to be both time-consuming and resource intensive. For example, finalizing the results from a single FreeSurfer recon-all execution can require over \SI{2}{\tera\byte} of RAM, while the developers recommend to use between \SI{8}{\giga\byte} to \SI{16}{\giga\byte} of RAM to run the pipeline\footnote{\href{https://surfer.nmr.mgh.harvard.edu/fswiki/SystemRequirements}{https://surfer.nmr.mgh.harvard.edu/fswiki/SystemRequirements}}. This significant difference in resource requirement creates a challenge to profile pipeline with long runtime, due to the potential difficulty to obtain access to compute nodes with enough RAM. One way to mitigate this challenge is to lower the sampling rate during the profiling. However, the sampling rate would have to be significantly reduced for the finalization stage to require a similar amount of RAM (\SI{16}{\giga\byte} or less), which would lead to a potential loss of information in the profiling results.
Because of these limitations, we were unable to finalize the results of the entire fMRIPrep pipeline and we only reported results for the sub-pipelines.

\section{Conclusion}
In this paper, we profiled several MRI pre-processing pipelines used in the popular fMRIPrep tool: ANTs brainExtraction, ANTs registrationSyN, FSL FAST, FSL FLIRT, FSL MCFLIRT, and FreeSurfer recon-all. We observed that few functions contribute to a large amount of CPU time which offers opportunities to optimize a practical amount of functions to potentially obtain significant speed-up values. We found that most pipelines suffered from memory bound, and that linear interpolation was the primary time bottleneck. We discovered a bug in ITK which leads the implementation of ANTs registration in single-precision to use double precision instead. We also discovered a potential bug in FreeSurfer recon-all which limits the benefit from multi-threading with OpenMP. Lastly, we discussed the challenge of profiling long-running pipelines with VTune, due to computational resource requirements.

The dataset we chose contains wide range of age and equal distribution of sex. However, it only contains healthy subjects. It would be interesting to study the impact of image quality and pathologies on performance. Moreover, our profiling with VTune did not account for the spatiality (e.g. for loop) or temporality (e.g. start vs. end of convergence) of the functions. This information could provide additional insight for optimization. Future work could extend the depth of the analysis by including this information.

Future optimization efforts could focus on reduced precision techniques, to reduce memory bound and accelerate interpolation computation. We note that the use of reduced precision is not trivial, as seen with the ANTs bug. Therefore, careful attention would be required to find a balance between performance and accuracy. We also suggest that future work study the different configurations of OpenMP scheduling in FreeSurfer recon-all, to improve parallel efficiency. This would lead to faster executions time and higher CPU usage in HPC allocations.

We hope that this work serves as a reference for future work to optimize MRI pre-processing pipelines. 

\section{Data Availability}
\label{sec:data-availability}
The entire OpenNeuro ds004513 v1.0.2 dataset is freely available to download at:
\\\href{https://openneuro.org/datasets/ds004513/versions/1.0.2}{https://openneuro.org/datasets/ds004513/versions/1.0.2}.
	
The container images for our profiling experiments are available on Docker Hub:
\begin{itemize}
	\item mathdugre/cmake:debug-info
	\item mathdugre/intel-compilers:debug-info
	\item mathdugre/afni:debug-info
	\item mathdugre/ants:debug-info
	\item mathdugre/fsl:debug-info
	\item mathdugre/freesurfer:debug-info
	\item mathdugre/fmriprep:debug-info
\end{itemize}
	
Our profiling results from VTune are available at:
\\\href{https://doi.org/10.5281/zenodo.10987491}{https://doi.org/10.5281/zenodo.10987491}

\section{Code Availability}
\label{sec:code-availability}
The code to download the data, compile and profile the pipelines, and generate the figures is available at:
\\\href{https://github.com/mathdugre/mri-bottleneck}{https://github.com/mathdugre/mri-bottleneck}
													
\section*{Acknowledgement}
Mathieu Dugr\'e was jointly funded by the Natural Sciences and Engineering Research Council of Canada (NSERC) through the Postgraduate Scholarship-Doctoral (PGS-D) program and the Graduate Doctoral Fellowship Award from Concordia University. This work was partially funded by the Canada Research Chairs program. The computing infrastructure was provided by the Canada Foundation for Innovation. 
													
\section*{Conflict of Interests}
The authors report no conflict of interests.
													
\bibliographystyle{IEEEtran}
\bibliography{paper}
													
\newpage
\onecolumn
\section*{Supplementary Materials}
\begin{figure*}[ht]
	\noindent
	\begin{minipage}
		{0.5\textwidth}
		\begin{align*}
			C & = CPU\_CLK\_UNHALTED.THREAD                \\
			S & = CYCLE\_ACTIVITY.STALLS\_LDM\_PENDING     \\
			O & = CYCLE\_ACTIVITY.STALLS\_L1D\_PENDING     \\
			T & = CYCLE\_ACTIVITY.STALLS\_L2\_PENDING      \\
			W & = MEM\_L3\_WEIGHT                          \\
			H & = MEM\_LOAD\_UOPS\_RETIRED.LLC\_HIT        \\
			R & = MEM\_LOAD\_UOPS\_MISC\_RETIRED.LLC\_MISS \\
			M & = \frac{W \times R}{H + R}                 \\
			N & = \frac{H}{H + W \times R}                 
		\end{align*}
	\end{minipage}
	\begin{minipage}
		{0.5\textwidth}
		\begin{align}
			\%~Memory~Bound & = \frac{S}{C} \times 100          \\
			\%~L1~Bound     & = \frac{S - O}{C} \times 100      \\
			\%~L2~Bound     & = \frac{O - T}{C} \times 100      \\
			\%~L3~Bound     & = \frac{T \times N}{C} \times 100 \\
			\%~DRAM~Bound   & = \frac{T \times M}{C} \times 100 
		\end{align}
	\end{minipage}
	\caption{Intel performance monitoring events and derived memory metrics}
	\label{fig:memory-metrics}
\end{figure*}

\label{sec:supplementary}

\csvnames{csvcol}{2=\module, 3=\func, 4=\mean, 5=\std}
\csvstyle{hotspot}{
	tabular = |r|l|l|c|,
	table head = \hline & Module & Function & CPU Time (mean$\pm$std)\\\hline\hline,
	late after line = \\\hline,
	respect all,
	csvcol
}

\begin{table}[ht]
	\centering
	\resizebox*{0.66\columnwidth}{!}{
		\begin{filecontents*}{tables/hotspots-32threads-freesurfer-reconall.csv}
,module_short,func_short,mean_cpu,std_cpu
1,libiomp5.so,__kmp_fork_barrier,134063.89932599998,12810.813820159035
2,libiomp5.so,omp_set_lock,34815.55207375,3599.58898506832
\end{filecontents*}
\csvreader[hotspot]{tables/hotspots-32threads-freesurfer-reconall.csv}{}
		{\thecsvrow & \module & \func & \tablenum[round-precision=2, round-mode=places]{\mean}$\pm$\tablenum[round-precision=2, round-mode=places]{\std}}
	}
	\caption{FreeSurfer recon-all (32 threads): Top functions accouting for 80\% of the pipeline makespan.}
	\label{extab:hotspots-32threads-freesurfer-reconall}
\end{table}

\end{document}